# Crypto-ransomware detection using machine learning models in file-sharing network scenario with encrypted traffic

Eduardo Berrueta [a], Daniel Morato [a,b,1], Eduardo Magaña [a,b], Mikel Izal [a,b]

[a] *Public University of Navarre, Department of Electrical, Electronic and Communications Engineering, Campus Arrosadia, 31006 Pamplona, Spain*
[b] *Institute of Smart Cities, calle Tajonar 22, 31006 Pamplona, Spain*

**Abstract**

Ransomware is considered as a significant threat for most enterprises since the past few years. In scenarios wherein users can access all files on a shared server, one infected host can lock the access to all shared files. We propose a tool to detect ransomware infection based on file-sharing traffic analysis. The tool monitors the traffic exchanged between the clients and the file servers and using machine learning techniques it searches for patterns in the traffic that betray ransomware actions while reading and overwriting files. The proposal is designed to work for clear text and for encrypted file-sharing protocols. We compare three machine learning models and choose the best for validation. We train and test the detection model using more than 70 ransomware binaries from 26 different strains and more than 2500 hours of 'not infected' traffic from real users. The results reveal that the proposed tool can detect all ransomware binaries, including those not used in training phase (unseen). This paper provides a validation of the algorithm by studying the false positive rate and the amount of information from user files that the ransomware could encrypt before being detected.

## 1. Introduction

Crypto-ransomware is a type of malware that extorts computer users by encrypting their files and requesting a ransom to recover the file content. In 2016, EUROPOL declared that ransomware was 'the most prominent malware threat […] for citizens and enterprises alike' [1].

Since 2018, crypto-ransomware attacks have been directed at companies in areas such as manufacturing, transportation, telecommunication, finance, public law enforcement, and health services [2] [3]. This is done because of the high economic profits that malware developers can gain from each infection. According to [4], 51% of the enterprises were attacked by ransomware in 2019.

In corporate environments, files are stored on shared networked volumes instead of users' local machines. This architecture facilitates implementation of backup policies, sharing capabilities, and security measures. In a volume-shared scenario, a single infected computer can encrypt all the files it has access to, thereby creating a highly compromised environment. An independent study of 5,000 IT managers across 26 countries [4] revealed that 65% of ransomware victims lost their data in network-shared volumes.

The proliferation of different strains of ransomware has given impetus to the development of detection tools focused on this type of malware. In a previous survey [5], we analysed more than 50 different tools, mainly from academia, and some from security companies. The most effective tools are based on monitoring disk-access activities [6] [7]. In a network file-sharing scenario, disk-access information can be obtained from network traffic. This facilitates tool deployment because herein, the tool is not required to run on every user host but can monitor file-sharing traffic from a network switch near the document repository.

---

1This is to indicate the corresponding author.
Email address: daniel.morato@unavarra.es

Traditionally, information contained in network file-access protocol messages is circulated in the clear on local area networks (LANs). However, owing to the popularization of public internet cloud services and increasing importance of confidentiality in network transactions, these protocols are evolving into their encrypted versions. Therefore, nowadays, a traffic monitor cannot obtain detailed information about the disk-access activities, and the tools based on such information do not work as desired. Until today, no tool is capable of ransomware detection based on encrypted network file-sharing traffic [5].

The aim of this study is to detect ransomware in action (i.e. during the encryption of network-shared files) by analysing features extracted from the encrypted network traffic. The large number of features offered by this traffic requires an analysis tool capable of detecting patterns in complex structures. In this study, we train and test three machine learning (ML) models. Using deep learning and an adequate set of features, the accuracy of detecting an active ransomware in 30 s reaches 99.8%, with a false positive rate of 0.004%.

The main contributions of this paper are:

- Presents a ransomware detection tool based on the analysis of encrypted network traffic in file-sharing scenarios. To the best of our knowledge, no previous proposal has targeted this specific scenario with encrypted traffic.
- Describes the processes of feature extraction, time-sample filtering, and parameter tuning for building an ML model capable of detecting ransomware in action, based only on the traffic exchange between the infected host and the file server. To make this study reproducible, we uploaded all the datasets and the optimized and trained ML model to a public repository [8].
- Compares the results of three different ML algorithms. The training and testing phases use more than 50 h of ransomware traffic from 73 different ransomware binaries and 50 h of real user traffic. The validation is based on ransomware strains not used in the training process (unseen binaries) and 2,477 hours of real user traffic.

The remainder of this paper is structured as follows: Section 2 summarises the literature on ransomware detection and highlights the unsolved problems tackled by the present proposals. Section 3 provides a detailed explanation of the scenario and the methodology, including the analysis of the datasets used for training, testing, and validating the models. Additionally, Section 3 describes the feature and time-sample selection and the metrics that will be used to evaluate the quality of the proposal. Section 4 presents the results of training, optimizing, and evaluating the ML models, selecting the one with highest accuracy and showing the existent trade-offs in its design. Section 5 provides a comparison of the results with those provided by other tools in the literature. Finally, Section 6 states the conclusions.

## 2. Background and related work

In a recent survey published in 2019 [5] we analysed more than 50 different ransomware detection tools. A few tools use traditional antivirus techniques based on the static analysis of program binaries before they are run [9] [10]. However, these techniques are prone to false negatives, so they are being substituted by solutions where indicators are extracted from monitoring the actions taken by any program running at the user's host. The tools use these indicators to classify running programs into benign or malign categories. The monitored indicators and the analysis functions are the main differences among the proposals.

The indicators describing the activities being taken by the program are most often extracted locally to the host where the program is running [6] [7] [11]. Some detection tools add information obtained from network traffic to the aforementioned indicators [12] or they work solely on the base of network metrics [13] [14] [15].

Detection tools based on local activities at the user's host monitor primarily disk access operations. Ransomware action can be detected on the base of its disk activity reading and writing files, as well as on the encrypted content of the written data. All this information is easily obtained by intercepting disk access input/output (I/O) operations [11] [16]. Complex detection tools add information such as the function calls to external libraries (searching for encryption libraries) [7] [17] or the directories in which the read or write operations are performed [6].

Pervious detection tools based on network traffic monitor domain name resolution requests, searching for patterns characteristic of dynamic name generation algorithms [13] [18] or they monitor the TCP connections to certain servers identified as malicious [12] [19].

The analysis function takes the indicators and uses them to accomplish the classification of the suspect program. The complexity of the analysis function varies in the literature. In some cases it only establishes thresholds to the measured metrics [20], while most often a combined metric from a large number of indicators is built [11]. In recent years, machine learning techniques have gained popularity thanks to their ease of use and their capacity of searching for patterns in a large number of features [7] [10] [12] [17] [21].

Since the publication of the survey [5], we have witnessed an increasing interest in ransomware detection techniques. This is the result of the high impact of ransomware in everyday activities and the risks affecting home users and enterprises. The new research proposals and industrial developments elaborate on the previously mentioned categories, based on the input data and the analysis algorithms.

Recent publications such as [22] [23] [24] [25] use ML models, taking as inputs features monitoring data from program actions at the user's host. Tools such as [26] and [27] are based on monitoring function calls and general network traffic, while the authors in [15] analysed the patterns in DNS requests, looking for suspicious characteristics. These new solutions usually compare some ML models, choosing the one with best results [26] [28], but sometimes the authors select the machine learning model without any comparison and try to improve its effectiveness by parameter tuning of the algorithm [25] or by changing the feature extraction process [29].

All the above-mentioned literature is focused on the traditional scenario of ransomware that encrypts local files to the user's computer. Although this is an omnipresent scenario, it is no longer where critical documents are stored in corporate environments. Medium-size and large companies use a network-centric architecture, where critical documents are stored in network-shared volumes. Most of these proposals can be applied to a file-sharing scenario because remote-access protocols offer an interface to the files using the local filesystem, therefore the system calls to access these drives can be intercepted at the host client. However, they require anti-ransomware software deployed in every user host.

Taking into account the specifics of a file-sharing network scenario, we envision a better approach to ransomware detection based on the analysis of the traffic between the users' computers and the file servers. This specific architecture has only been studied previously in REDFISH [30]. However, REDFISH analyses the protocol messages in the file-sharing protocol. It requires the metadata about the disk I/O operations – contained in the file-sharing protocol messages – therefore it can only be used when the file-sharing protocol does not send encrypted messages. Please beware of the distinction between encryption at the protocol level and encryption at the application level. Even if the application is ransomware trying to overwrite a file with encrypted content, a clear-text file sharing protocol will send protocol messages where the type of operation, the file path, the byte position in the path, will all be on-the-clear, even though the content written were encrypted. REDFISH required clear-text protocol messages; however, this is not the case in new file-sharing protocols that use encrypted transport.

To the best of our knowledge, no previous work in the literature or industrial tool has solved the problem of detecting ransomware activity on the base of encrypted file-sharing traffic. We believe it is an interesting scenario that can take advantage of characteristics such as the better scalability due to not requiring installing monitoring software in any host or file-sharing server and, being completely off-path, it does not interfere with user actions. We take from the recent literature the approach of using ML techniques, which are adequate to capture the large variability of behaviours in the actions from benign and malign programs.

## 3. Scenario and methodology

Servers with shared volumes are growing in popularity in home environments – this is owing to the increase in the number of devices connected to the same home network and the advantages for example for multimedia content sharing between devices. However, we focus this work on large corporate networks where shared volumes store files that can be accessed by multiple users, facilitating collaborative work, mobility and the implementation of backup policies.

Figure 1 illustrates a population of users in a corporate local area network (LAN) accessing files from a common server. Ransomware running at one of the hosts reads large amounts of data from the files stored in the server and writes the encrypted version of those files in the same server. Ransomware detection can be accomplished by detecting these read and write actions, as well as the delete or rename actions that the ransomware performs during its progress. A single infected host can pose significant danger in this scenario because it could encrypt all the shared files (or at least those it has access to).

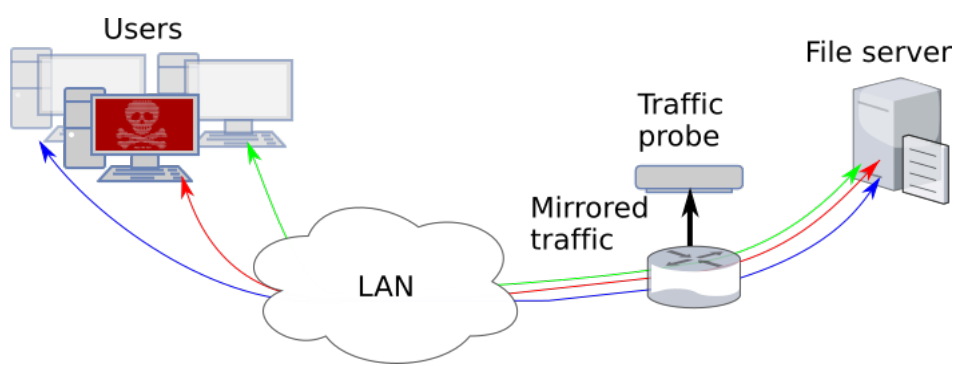

**Figure 1: Monitoring scenario with ransomware detection tool based on network traffic.**

A ransomware detection tool based on monitoring network traffic can be installed either on every user host, on the server, or in the network path between the server and users. Installing the tool on every user host adds significant complexity to the deployment and could reduce computer responsiveness. A simpler solution is to monitor the traffic at the file server; however, the tasks of network monitoring and ransomware detection can impact the server performance and consequently, all users. Figure 1 shows the third option – a network traffic probe monitoring messages between users and file server. This deployment option has been used in previous proposals [30]. Most enterprise-grade ethernet switches offer the capability of port mirroring, i.e., duplicating the traffic from one port to another. A device with the detection tool installed on it, can be connected to a port-mirror ethernet switch port. It can capture and analyse the traffic without any effect on network latency, response time, or computer responsiveness because the probe is not in the traffic path but monitoring a copy of the traffic. Using commodity hardware, traffic rates in the order of tens of gigabits per second can be captured and processed [31].

Recent protocols in network file-sharing scenarios are commonly transported over TCP/IP. The most commonly used protocol, both in enterprise and home deployments, is server message block (SMB), particularly, its second and third versions (SMBv2 and SMBv3, respectively). Despite the availability of other file-sharing protocols such as network file system (NFS) [32] and Apple filing protocol (AFP) [33], the extended use of the Windows operating system (OS) in corporate environments makes SMB the most popular protocol because it is the default file-sharing protocol used in this OS.

Network traffic offers a plethora of metrics: number of TCP connections, bytes transferred, sequence of messages between client and server, packet sizes, inter-packet times, inactivity times, connection durations, and sequences (in time) of any of these, to name a few. We resort to ML techniques, which have been validated in previous works as adequate tools in this type of scenario [26] [29]. ML algorithms can analyse these complex data structures, and once they are trained with a complete dataset, they can generalise to different input data. The most adequate set of features used in training the ML model depends on the characteristics of file-sharing traffic. In the following sections we describe the network protocol scenario, the methodology for constructing the feature dataset we used in training and testing, the models that will be evaluated, and the metrics used for this task.

## 3.1 Protocol behaviour

In this study, we consider a network file-sharing scenario where the server and clients communicate using the SMB protocol. In [30], we considered such a scenario, although only for version 2 of the protocol (version 1 has been deprecated since 2014 [34]). We extended the results to SMBv3, which implies significant changes in network traffic. SMBv3 is the default and recommended version since Windows 8 and Windows Server 2012 [35] . This version introduced message encryption and some other changes to the original protocol, making it more secure. The added encryption layer makes it impossible for traffic analysis tools to distinguish messages and different user activities. This version is expected to become the most popularly deployed version of the SMB protocol with enterprises migrating from deprecated versions of the Windows OS [36].

SMB is a request-response protocol transported over TCP using a single connection to a well-known server port (value 139 or 445), and with 19 different commands in its second and third versions. Each command corresponds to one action that the client issues over a server file or directory. These commands can be considered similar to the input/output operations (I/O ops) performed locally at the user's computer. Moreover, our previous investigation has revealed that it is possible to detect ransomware in action, based on the commands from the SMBv2 protocol [30].

Owing to the encryption of data over TCP in SMBv3, the command type and parameters cannot be identified. Tools based on the analysis of these commands cannot successfully detect ransomware in the encrypted version of SMB. Although it is impossible to view the client operations, some features can be extracted from the encrypted network traffic, which could allow accurate identification of the ransomware actions. Although this study is focused on file sharing using the SMB protocol, the procedure employed in it is extensible to other request-response file-access protocols such as NFS, even with encrypted traffic.

Figure 2 illustrates a typical SMBv3 packet sequence, wherein the user opens, reads, and closes a file on the server. Figure 2(a) illustrates the SMB packets captured from the network, and Figure 2(b) illustrates the unencrypted commands. By focusing on these unencrypted commands, one can view all the operations performed by the client, including the file metadata (name and file size) and the number of bytes read. This information is encrypted in the network (Figure 2(a)), and it is not possible to know the operations performed by the client or the file that is being worked on. It is possible to extract other features from the encrypted traffic, based on the time between packets or packet size.

*3.2 Dataset*

Two types of samples are used in this study: (i) those from traffic captured while ransomware was encrypting network-shared files ('infected') and (ii) those from office users running benign applications and accessing shared files ('not infected').

The traffic traces for the 'infected' case were obtained from a repository we built and shared publicly in [37]. This repository comprises traffic traces from more than 70 ransomware programs grouped in 26 strains. Each of these 70 binaries were executed more than once, generating 150 traffic traces in total. They were captured while the malware was encrypting a large file set shared by an SMB server. We obtained more than 50 h of ransomware activity from these traffic traces.

From these ransomware traffic traces, we separated the five most recent ransomware families. They are not included in the training set but further used as examples to check whether they are accurately detected or not, even when they are not part of the learning process (we name them 'unseen' samples). These five families are: bitPaymer (appearing in November 2019) [38], Shade (November 2019) [39], Sodinokibi (March 2020) [40], Phobos (May 2019) [41] and Stop (February 2020) [42].

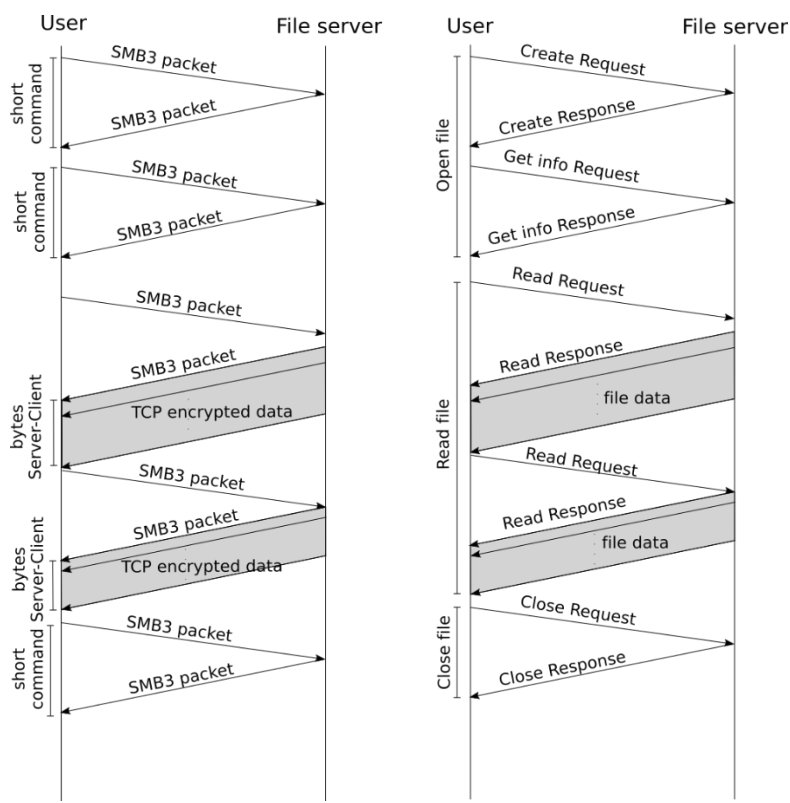

**Figure 2: Packet sequence in (a) encrypted version and (b) non-encrypted version of SMBv3.**

The 'not infected' samples were obtained from network traffic traces captured in a campus LAN, wherein users access files from a shared server. We collected 2,527 h of 'not infected' traffic. This represents 316 intervals of 8 working hours. For each 8-hour connection, on average, 7,640 files were opened, and 91.4 and 294 MB data per connection were read and written, respectively.

From these three groups ('infected, 'not infected', and 'unseen'), we extracted samples to create three different datasets. The first: the training and testing dataset comprises all ransomware samples labelled as 'infected' and a selected set of 50 h of user samples labelled as 'not infected'. We used 80% of these first dataset for training the algorithm and the remaining for testing it. During the testing phase, we measured accuracy, F-measure, precision, recall, and phi-coefficient of each model [15]. Subsequently, we selected the most effective algorithm.

The second dataset comprises 2,477 h of user activity labelled as 'not infected' and not included in the training and testing dataset. This dataset is used to test the false positive rate, i.e., measure situations wherein ransomware is falsely detected. This rate determines the usability of the tool because a large number of false positives render the tool unusable for a real environment.

Finally, the 'unseen' dataset comprises samples extracted from the newest ransomware traffic traces. Using this dataset, the capability of the tool to detect unseen ransomware binaries is measured. Furthermore, we measure the number of files and bytes that the ransomware encrypts before detection.

## 3.3 Feature extraction

The following is the fundamental characteristic of all ransomware programs that has remained immutable since the appearance of the first strains of crypto-ransomware in 2014 – they always read and write large amounts of bytes for encrypting user files. While some ransomware strains overwrite the content of the original files, others create new encrypted files before the deletion of the original. In both cases, reading the original and writing the new encrypted one must be performed by ransomware before trying to extort the users. Ransomware tries to complete this process as fast as possible to avoid being detected by the user when she tries to open a file and discovers that it is unreadable. However, the speed of these actions depends on the complexity of the encryption algorithm, the efficiency of its implementation, the hard disk speed or the CPU power. Several tools in the literature base their detection on the observation of these read and write operations [5].

Owing to the encryption of traffic content in recent file-sharing protocols, a significant amount of metadata about these actions remains hidden from network traffic monitors. These monitors can only record packet arrival times, sizes, and direction (from client to server and vice versa). Based on this information, the network probe can only guess the file-access actions performed by the user. These actions could involve opening a file, closing it, obtaining information about it, changing its metadata, writing content to it, or reading content from it. Based on the aforementioned limitation, we define the following three actions:

- Bytes are being written (Figure 3(a)): We consider a write operation when there is a one-packet response for a large (more than one packet) request.
- Bytes are being read (Figure 3(b)): We consider a read operation when there is a one-packet request and a large response (more than one packet). In Figure 2(a), we term these bytes: 'bytes Server-Client'.
- Control or short commands (Figure 3(c)): These commands include operations such as delete, rename, open, and close file. They do not require a large amount of data describing the command; they can usually fit into a single-packet request from client to server and a single-packet response from server to client. Beware that short read or write actions whose data fits in a single packet are indistinguishable from control commands, however, the operations system tends to batch disk access operations to optimize data flow, making this event unlikely.

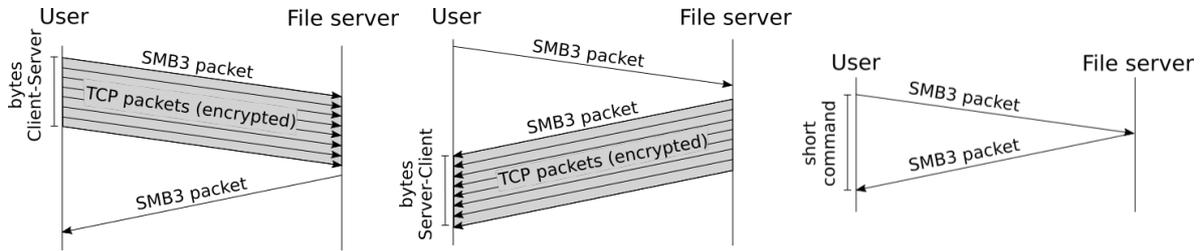

Figure 3: Bytes being written (a) read (b) and short command (c) between client and server.

Ransomware tries to encrypt as many files as possible before being detected. Therefore, it performs frequent file-access operations, compared to a typical office program. However, in some user behaviours, the number of bytes read and written can be equal or higher than that by ransomware during a short period of time. One example of these intense behaviours is the duplication of data in the server.

Figure 4 presents the bytes (a) read and (b) written per second by a ransomware (Cerber) encrypting a shared directory, and by a user while duplicating all the files in the directory. Herein, the user activity is greater than that of the ransomware because the user does not suffer the burden of encrypting the files. It is not easy to distinguish the two cases by only considering the bytes read and written, the inter-arrival packet times or the packet sizes. To this end, we need a feature that can differentiate these extreme cases of similar activity. Figure 5 presents the number of short commands per second for the same traffic traces. It can be observed that this number is significantly lower for directory duplication. In-depth study of ransomware behaviour reveals that when ransomware encrypts files, it must delete the original ones and sometimes create extra files in each directory with payment and decryption instructions for the victim. These actions lead to the difference in the number of short commands and enables us to differentiate both cases.

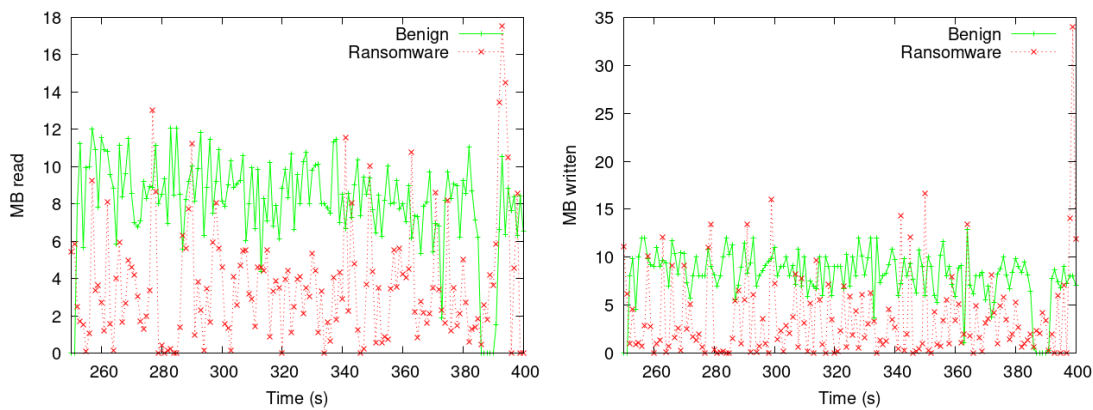

Figure 4: Bytes (a) read and (b) written per second by ransomware and user.

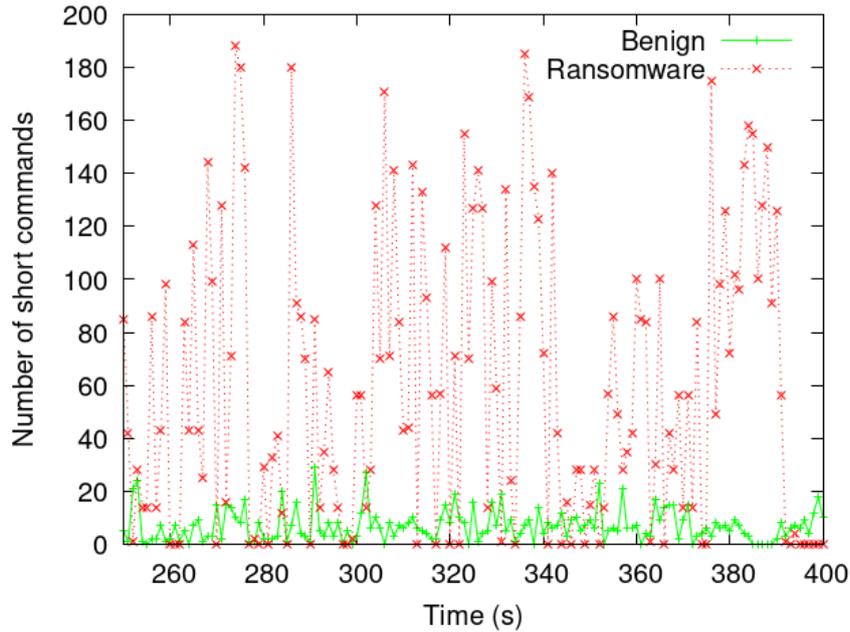

**Figure 5: Short commands per second performed by user and ransomware.**

Consequently, similar to the ransomware detection techniques run locally on the infected host, ransomware action can be recognised from network traffic based on the traffic pattern between the client and server. When network traffic comprises clear-text the ransomware detection tool can also use the file-access action types to assist in differentiating ransomware actions from benign applications [30]. The analysis tool knows the commands being used (changing file names, deleting or overwriting files) and it can also measure the higher entropy in written encrypted file content. In scenarios where all disk access commands are encrypted, these metadata about user actions are not available, however the aforementioned new feature, based on the number of control or short commands, can still be utilized as a differentiator between both types of actions.

All these features cannot be extracted for a single packet – they are the result of traffic accumulated during a (preferably) short period of time. The larger is the analysis window, the easier it is to distinguish ransomware from benign applications, owing to the larger variability in disk-access patterns in the latter. However, the larger the period of time to detect the ransomware, the larger the number of files lost before the ransomware can be blocked.

Analysis of the abovementioned traffic features (number of bytes read/written or control commands) is performed for each TCP connection between one client and the server. However, analysis over a short period of time, e.g., one second can result in several user behaviours that cannot be differentiated from ransomware. Therefore, larger time windows that contain clear behavioural patterns are required. Consequently, although we measure the traffic features in per-second time intervals, we introduce a temporal window of T seconds to create complete time-samples for the learning process. Hereafter, the term 'sample' means the time windows composed by 3*T features that are input into the classification model. These samples should not be confused with the ransomware executables that are sometimes called 'samples' in the literature. We refer to the latter as 'ransomware binaries'. Figure 6 illustrates an example for T=10 s, where N=3*10=30 features are present in each sample. For each 1-second interval, the traffic probe computes the following.

- Total number of short commands where the response is contained inside the window.
- Total number of data (TCP bytes) in the packets sent from the server to client that are not part of short commands
- Total number of data (TCP bytes) in the packets sent from client to server that are not part of short commands.

These features represent the control commands, read actions, and write actions, respectively. The complete sample for the machine learning model comprises these three values for every second in the time window of T seconds.

The next sample is created with traffic out of the window from the previous one. We can slide the time window a small amount of time or as much as the window length T, creating samples without shared data. In the example of Figure 6, the next sample without shared data will include the features between the 10$^{th}$ and the 20$^{th}$ second; therefore, we generate one sample every T seconds. A new sample with shared data could be built by sliding the windows for example only 1 or 2 seconds, creating a new set with data from the 1$^{st}$ to the 11$^{th}$ second or from the 2$^{nd}$ to the 12$^{th}$ second respectively.

Parameter T and the sliding window step are tuneable. The detection accuracy and data loss depend on them. With an increasing T, the samples comprise an increasing number of time-intervals, and the algorithm can learn more complex relations between features. However, detection of ransomware would require more time (because it needs at least T seconds) and the user will lose more data in the process. With a small sliding step, the new sample does not contain much new information, however the detection algorithm could react earlier to abrupt changes in the traffic pattern. We will evaluate the effect of both strategies in the optimization process.

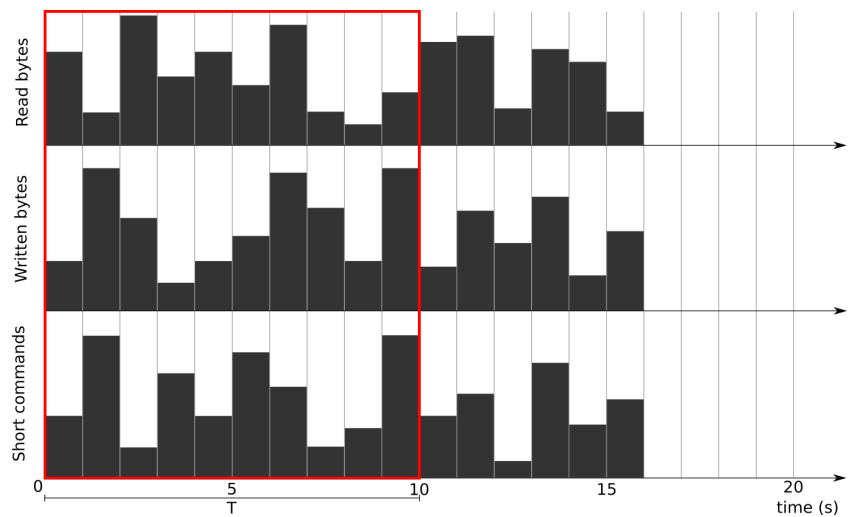

**Figure 6: Example of feature extraction in T seconds.**

## 3.4 Sample filtering

ML models require significant and adequate sets of learning samples for the behaviours that they must detect [43]. The different activities (from benign applications and malware) must be accounted for without over-representation of some of them, which could result in ignoring significant patterns.

Network file-sharing traffic from office users is highly intermittent, with large thinking times, for which no significant amount of traffic occurs. Samples from benign users must be selected by considering the higher popularity of time periods for which no traffic exists and being especially cautious by providing enough learning samples where user actions create network traffic, such that it is not recognised as ransomware traffic.

We studied each of the three features while comparing 'not infected' and 'infected' samples. In Figure 7, the complementary cumulative distribution functions for the bytes written per second in an 'infected' trace and a 'not infected' one are plotted. For ransomware samples, only 20% 1-second intervals contain no written bytes, whereas for user traffic, more than 99% 1-second intervals are inactive intervals in terms of written bytes. A similar situation is revealed for the bytes read. The training of the ML algorithm requires a significant number of 'not infected' samples where the user is active. This number is not proportional to their presence in relation to non-active intervals. Otherwise, the learning process could ignore rare high-activity benign users and generalise that high activity is always indicative of ransomware action [43].

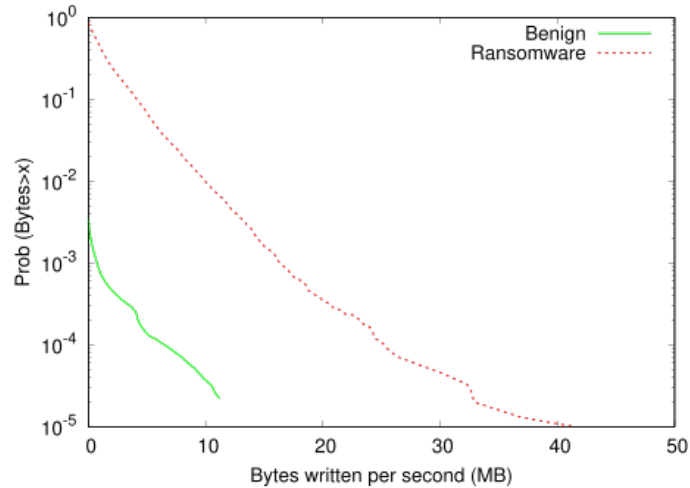

**Figure 7: Complementary cumulative distribution function for the bytes written per second.**

In addition to the randomly selected samples, we considered all time-samples with one or more 1-second intervals with more than 5 MB data read or written. Although some 'not infected' samples could contain time intervals with large number of file-access operations, ransomware samples typically contain longer high-activity intervals (see Figure 4 and Figure 5).

Additionally, we included in the learning dataset all the samples from the 'not infected' users with any 1-second interval containing more than 100 short commands. Therefore, the algorithm is expected to learn that clean samples can exist where high activity is possible (in terms of control traffic).

In the learning process, all the ransomware samples were used, ignoring only the inactivity time before the first file was opened for encryption and after the last file was altered.

## 3.5 ML models and evaluation metrics

ML models can find complex patterns in a large number of features. The solution presented in this paper analyses three ML models: decision trees (DTs), three ensembles (TEs), and neural networks (NNs). DTs are the simplest of these three models; however, they are susceptible to over-fitting. They have been used for ransomware detection in research articles such as [15] [22] [44]. TEs combine some DTs and are capable of finding more complex relations between features, although at the expense of a higher model complexity. The authors of [15] and [24] analysed the capability of TEs to detect ransomware, achieving high detection rates. Finally, NNs, owing to their flexibility, are popular models in the literature [9] [15] [25] [45] [46]. NNs are more complex than DTs or TEs; however, model computation is only required once every T seconds (see Section 3.3) which does not impose a critical speed requirement.

These three models were trained and tested using bigML [47] with the dataset described in Section 3.2. The best model was selected based on several binary classification metrics. These metrics can be derived from the confusion matrix, and they are defined in Equation 1-5, where TP means 'true positive', FP means 'false positive', TN means 'true negative', and FN 'false negative'.

$$\text{Accuracy} = \frac{TP + TN}{TP + FP + TN + FN}$$

**Equation 1**

$$\text{F} - \text{measure} = \frac{2 * Precision * Recall}{Precision + Recall}$$

**Equation 2**

$$\text{Precision} = \frac{TP}{TP + FP}$$

**Equation 3**

$$\text{Recall} = \frac{TP}{TP + FN}$$

**Equation 4**

$$\text{Phi coefficient} = \frac{TP * TN - FP * FN}{\sqrt{(TP + FP) * (TP + FN) * (TN + FP) * (TN + FN)}}$$

**Equation 5**

For computing these metrics, each sample contains N features, covering a time interval of T seconds.

In a ransomware detection technique based on the dynamic analysis of actions taken by the suspect program, the number of false positives has a more significant effect in the usability of the tool than the number of false negatives. One false negative means that the ransomware has not been detected in the first T seconds, but it may be detected in the next time window. However, a high number of false positives could reduce user confidence in the detection system, thereby increasing the possibility of the user ignoring a true alarm. Consequently, in the metrics presented, precision is more relevant than recall.

To reduce the number of false alarms and maximize true detections, we tune the value of the time windows length parameter (T) in each model (see Section 4.1) The model with the best results is consequently optimized and validated in Section 4.2. However, on the final evaluation we have not considered only the traditional metrics for a binary classifier, but we have added the following scenario-specific metrics:

- Time for ransomware detection: The time that the tool takes to detect ransomware. It depends on the window length parameter.
- Data lost before ransomware detection: The amount of data (measured in megabytes) lost due to encryption. We account for the whole file size even if the ransomware encrypts only part of the file before it is detected. It is a pessimistic metric, but it is safer to assume that the file may not be recoverable even if only part of it was encrypted.
- Number of working days until a false positive: It is the number of 8-hours working days that it is expected to be without a false positive. It is computed based on a dataset containing 309 8-hours working days (those not used in the training phase).

## 4. Results of model selection, validation and optimization

We compared three different ML models using the techniques for optimization offered by BigML. These are centred on ML metrics such as those described in Section 3.5. The best model obtained from this broad comparison was studied in detail for specific metrics in our scenario, such as the number of bytes lost or the time until ransomware detection.

In this section we present the results from each of these steps.

## 4.1 ML model selection

We considered six values of interval T, from 10 to 60 s. For each value we prepared the training and testing dataset, trained each of these ML models and computed the classification metrics.

The metrics computed for the six values of T are shown in Table 1 Table 1: Test results for DTs.(Decision Trees), Table 2 (Tree Ensembles) and Table 3 (Neural Networks). As T increases, the results are expected to improve because a larger number of features allows the model to learn more complex relations. However, using large values of T causes the model to require more time to raise an alarm, allowing the ransomware to encrypt more files before it is detected. This effect on the detection time is studied in-depth in Section 4.2, after the best model is selected.

All the models achieve high accuracy results (98% or higher). However, the simpler the model is, the less it takes advantage of the larger number of features. By using NNs, both accuracy and precision increase as T becomes larger (larger number of features). By using DTs, both accuracy and precision decrease when the temporal window, T, increases.

| T. samples (s) | Accuracy (%) | F-measure | Precision (%) | Recall (%) | Phi-coefficient |
|---|---|---|---|---|---|
| 10 | 98.9 | 0.9865 | 98.5 | 98.8 | 0.9779 |
| 20 | 98.8 | 0.9815 | 98.1 | 98.1 | 0.9725 |
| 30 | 98.7 | 0.9786 | 98.4 | 97.3 | 0.9697 |
| 40 | 98.8 | 0.9783 | 98.6 | 97.1 | 0.9704 |
| 50 | 98.8 | 0.9754 | 97.9 | 97.2 | 0.9671 |
| 60 | 98.8 | 0.9739 | 98.5 | 96.3 | 0.9661 |

**Table 1: Test results for DTs.**

| T. samples (s) | Accuracy (%) | F-measure | Precision (%) | Recall (%) | Phi-coefficient |
|---|---|---|---|---|---|
| 10 | 99.5 | 0.9935 | 99.2 | 99.5 | 0.9894 |
| 20 | 99.4 | 0.9913 | 99.1 | 99.2 | 0.9871 |
| 30 | 99.5 | 0.9914 | 99.5 | 98.8 | 0.9878 |
| 40 | 99.6 | 0.9930 | 99.8 | 98.8 | 0.9904 |
| 50 | 99.6 | 0.9913 | 99.9 | 98.4 | 0.9885 |
| 60 | 99.6 | 0.9909 | 99.6 | 98.6 | 0.9882 |

**Table 2: Test results for TEs.**

| No. of hidden layers | T. samples (s) | Accuracy (%) | F-measure | Precision (%) | Recall (%) | Phi-coefficient |
|---|---|---|---|---|---|---|
| 1 | 10 | 99.1 | 0.9886 | 99.0 | 98.7 | 0.9813 |
|   | 20 | 99.1 | 0.9868 | 99.2 | 98.2 | 0.9804 |
|   | 30 | 98.8 | 0.9789 | 98.9 | 96.9 | 0.9703 |
|   | 40 | 99.2 | 0.9854 | 98.7 | 98.4 | 0.9810 |
|   | 50 | 99.3 | 0.9860 | 99.3 | 97.9 | 0.9813 |
|   | 60 | 99.5 | 0.9896 | 99.4 | 98.6 | 0.9865 |
| 2 | 10 | 99.1 | 0.9887 | 98.9 | 98.8 | 0.9816 |
|   | 20 | 99.3 | 0.9888 | 99.2 | 98.6 | 0.9833 |
|   | 30 | 98.9 | 0.9807 | 98.4 | 97.8 | 0.9726 |
|   | 40 | 99.2 | 0.9849 | 98.6 | 98.4 | 0.9793 |
|   | 50 | 99.3 | 0.9853 | 99.5 | 97.6 | 0.9804 |
|   | 60 | 99.6 | 0.9910 | 99.1 | 99.1 | 0.9882 |
| 3 | 10 | 99.7 | 0.9959 | 99.7 | 99.4 | 0.9933 |
|   | 20 | 99.8 | 0.9966 | 99.5 | 99.8 | 0.9950 |
|   | 30 | 99.8 | 0.9971 | 99.6 | 99.8 | 0.9959 |
|   | 40 | 99.8 | 0.9962 | 99.6 | 99.7 | 0.9948 |
|   | 50 | 99.9 | 0.9974 | 99.7 | 99.7 | 0.9965 |
|   | 60 | 99.9 | 0.9987 | 99.7 | 100 | 0.9983 |

**Table 3: Test results for NNs with 1, 2 and 3 hidden layers.**

Figure 8 plots the six metrics for the three models, using T=20 s. To facilitate easier comparison, for NNs, we plotted only the results with three hidden layers, which offers better results than when using 1 or 2 layers. In Section 4.2, we present a detailed comparison of the other metrics for this model with different number of hidden layers.

DTs provide the worst results for all the metrics. The difference between DTs and the other models is maintained for other values of T – it is larger for the case of T=60 s (Table 1). DTs cannot take advantage of all the features offered, unlike TEs and NNs. Therefore, we discard the DT model and focus on TEs and NNs.

NNs provide the best results, with values greater than 99% for all the metrics. Moreover, the phi-coefficient for TEs has the lowest value in the set (98.7%). Although the differences in their case are not as large as in the case of DTs, NNs obtain higher values for all the metrics for most of the T values studied (10-60 s). In Section 4.2, we discuss the optimization of the NN model studying the effect of the changes in parameter T and the number of hidden layers on the false positive rate and the accuracy results on the unseen ransomware binaries.

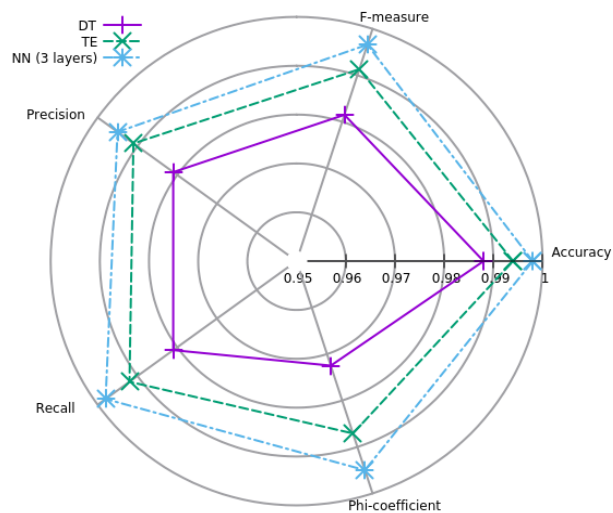

**Figure 8: Comparison of metrics for DT, TE, and NN with T=20s.**

## 4.2 Neural network model optimization

NNs provide the best results among the models studied; however, they have scope for optimization. We evaluated the true positive and false negative rates using the required time until ransomware is detected in the 'infected' dataset. We computed the average and maximum time required by the model to raise an alarm for the 150 ransomware traffic traces available. Each model requires to wait for at least T seconds to raise an alarm because T is the time needed to build a single sample. All NN models in this section can detect all the ransomware binaries in our dataset – this includes those not used in the training phase (unseen malware) – however, different configurations require different time to detection. Table 4 lists the results for NNs with one, two, and three hidden layers with different configurations of the time window.

The last column in Table 4 presents the maximum time required to detect a ransomware in the 'unseen' dataset. These strains were not used in the training phase, as explained in Section 3. This is the most significant result because it reveals the robustness of the system to the ever-changing environment of malware. All the models detect the unseen ransomware in the time required to build a sample (T seconds) except for two scenarios: (a) when T=10 s, there are some strains that require two consecutive samples (20 s) to be detected, and (b) for the 1-hidden layer model and T=60 s, that two samples may be required, raising the maximum detection time to 120 s.

| Hidden layers | T. samples (s) | False positives | | Average / maximum detection time to detect training binaries (s) | Maximum time to detect any unseen binary (s) |
| --- | --- | --- | --- | --- | --- | --- |
| | | Number | % | Average number of workdays until a false positive occurs | | |
| 1 | 10 | 207 | 0.0230 | 1.49 | 10.88/30 | 20 |
| | 20 | 25 | 0.0057 | 12.3 | 22.62/260 | 20 |
| | 30 | 9 | 0.0031 | 33.6 | 34.13/630 | 30 |
| | 40 | 6 | 0.0027 | 51.4 | 41.9/280 | 40 |
| | 50 | 4 | 0.0023 | 75.4 | 52.46/1050 | 50 |
| | 60 | 3 | 0.0021 | 99.2 | 62.17/540 | 120 |
| 2 | 10 | 386 | 0.0430 | 0.8 | 10.68/40 | 20 |
| | 20 | 76 | 0.0174 | 4 | 21.37/160 | 20 |
| | 30 | 26 | 0.0090 | 11.5 | 31.4/180 | 30 |
| | 40 | 11 | 0.0051 | 27.2 | 42.5/240 | 40 |
| | 50 | 9 | 0.0049 | 34.7 | 50.7/700 | 50 |
| | 60 | 6 | 0.0042 | 49.6 | 61.3/300 | 60 |
| 3 | 10 | 76 | 0.0080 | 4.07 | 10.2/20 | 20 |
| | 20 | 35 | 0.0080 | 8.84 | 20.2/40 | 20 |
| | 30 | 12 | 0.0041 | 25.8 | 30.2/60 | 30 |
| | 40 | 7 | 0.0032 | 43.8 | 40/40 | 40 |
| | 50 | 7 | 0.0040 | 43.8 | 50/50 | 50 |
| | 60 | 4 | 0.0028 | 74.8 | 60/60 | 60 |

**Table 4: Validation results for NNs with 1, 2, and 3 hidden layers.**

The maximum detection time is also high for the training dataset when using the 1 or 2 hidden layer models and T>10 s. The maximum detection time is 160 s when using 2 hidden layers and T=20 s and it is larger than 15 min (1050 s) when using only 1 hidden layer and T=50 s. This means that some ransomware strains can run for more than 15 min, encrypting user files, before they are detected. Considering only the tool effectiveness to detect ransomware in action we should select a 3-hidden layers model or a 1-hidden layer model but the last case only using T=10 s. The 3-hidden layers model is complex but has an adequate behaviour for any value of T between 10 and 60 s. The 1-hidden layer model is simpler, but it can take too long to detect some ransomware families, so it should only be used with T=10 s.

However, we cannot base model optimization only on detection effectiveness. The system will not be usable if it frequently raises an alarm when benign applications are accessing shared documents. A high number of false positives could disturb the user's workflow, dropping her confidence in the detection system and resulting in it being deactivated. It is very important to keep the number of false positives low for the tool to provide a satisfactory user experience. For the evaluation of these false positives, we used the real user samples labelled as 'not infected'. They contain more than 10 months of 8-hours working days of traffic. We counted the number of T-seconds long samples mistakenly classified as 'infected' for each value of T. Furthermore, we calculated the average number of 8-hour working days before raising a false alarm, which has an inverse relationship with the false positive rate. The results of false positives are in the third, fourth and fifth column in Table 4. The number of false positives gets reduced when increasing the number of layers or the time window T. This reduction results in larger periods of time before a false alarm is raised when only benign applications are used.

Looking at the average number of days until a false positive, the 1-hidden layer model obtains the best results for all configurations except for T=10 s. When T=10 s, the 1-hidden layer model offers an average of 1.49 days between false alarms; meanwhile, the 3-hidden layer model increases this value to 4.07 days. The best result is obtained using T=60 s and 1 hidden layer, where only one false positive is expected ever 99 8-hours working days. However, this behaviour is the result of also not detecting some ransomware strains early enough. The 1-hidden layer model is conservative in raising an alarm, providing good results of false positive rate at the expense of late ransomware detections. We have established that this model shows good detection results only when T=10 s. However, for such a low value of T the 1-hidden layer model does not provide the lowest false positive rate, but it is surpassed by the 3-hidden layers model. Therefore, we discard the 1-hidden layer model as the best option.

The 3-hidden layers model provides a false positive rate lower than the 2-hidden layers model for any configuration and it also has the best behaviour when T=10 s. This means that for low T, the 1-hidden layer model shows low maximum detection times, but it does not offer the best result in false positive rate. The best result is offered by the 3-hidden layers model, which also offered good detection results. Although the 3-hidden layers model requires more time for training and classifying each sample, the model is trained only once after its installation; therefore, the time utilized for training is not determinant. The NN receives one sample every T seconds, and NN models using hundreds of neurons can be evaluated in sub-second time on a single-core CPU; therefore, the classification time is also not a cause of concern.

Once we have selected the model with best results, we should select the best value of T. No detection can happen before a single sample is read, which requires the traffic in a time window T. The 3-hidden layers model offers detection in one time window except for T=10 s that requires a second sample for some ransomware strains, taking at most 20 s to detect the ransomware.

Figure 9 shows the number of bytes encrypted for each unseen binary before successful detection using the 3-hidden-layer model. Owing to the differences between the encryption processes of different ransomware binaries, the number of bytes encrypted can vary greatly among them. Some ransomware strains encrypt files in the alphabetic order, whereas others do it according to size [37]; this is the reason for the differences in the number of bytes encrypted.

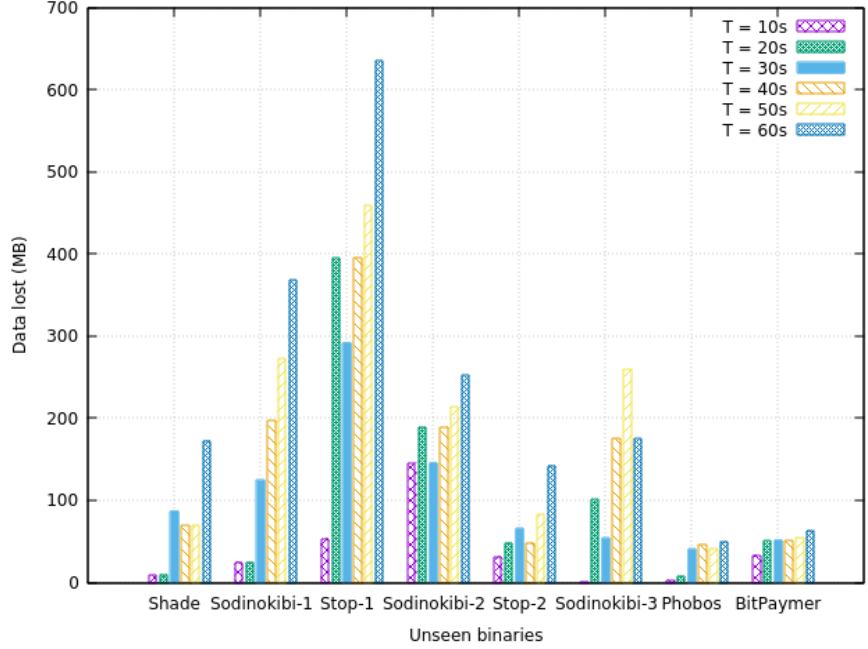

**Figure 9: Data in MBs lost for each unseen binary for different configurations of T (3-hidden layers NN).**

For T=60 s, the average number of bytes lost for unseen ransomware binaries is 232MB, whereas for T=10 s, it reduces to 37MB. Ransomware such as BitPaymer are considerable insensitive to parameter T, whereas others such as Stop-1 can vary from encrypting less than 100 MB of data when T=10 s to encrypting more than 6 times this value when T increases by 6 times (T=60 s). One of the objectives of the tool presented in this paper is quick detection of ransomware; therefore, we recommend that parameter T should not be greater than 30 s. This allows unseen binaries to encrypt an average of 113 MB data. By using T=30 s, a false positive rate of 0.0041% and one expected false alarm only after 25 working days are seen.

Depending on the effect of the alarm on the user, shorter values of T can be configured. If the alarm causes only a warning, it may be acceptable to have one false positive in 4 days (T=10 s). This would result in less data loss in case of ransomware infection. If the alarm is annoying to the user, larger values must be used for T. Parameter T is tuneable through the tool or network administrator, and it depends on the network, user, and server characteristics.

Once the time-window length T is decided we can evaluate different options in the process of sliding the time window to create the input samples from the network traffic. Table 5 shows the number of megabytes lost for each of the ransomware binaries in the 'unseen' set. Different values of the sliding step were evaluated, from 1 s (the minimum interval to extract the features) to 30 s (the selected window length). Table 6 contains the number of false positives for each of these configurations.

| Unseen binary | 1 s (MB) | 5 s (MB) | 10 s (MB) | 20 s (MB) | 30s (MB) |
|---|---|---|---|---|---|
| Shade | 75 | 40 | 82 | 33 | 87 |
| Sodinokibi-1 | 167 | 129 | 129 | 129 | 129 |
| Stop-1 | 351 | 365 | 398 | 478 | 328 |
| Sodinokibi-2 | 142 | 164 | 188 | 188 | 146 |
| Stop-2 | 60 | 59 | 68 | 68 | 68 |
| Sodinokibi-3 | 0.27 | 0.27 | 0.27 | 54 | 54 |
| Phobos | 12 | 13 | 13 | 14 | 41 |
| BitPaymer | 33 | 33 | 51 | 33 | 53 |
| Average (MB) | 105 | 100 | 116 | 124 | 113.25 |

**Table 5: MB lost before detection of each unseen binary for different sliding steps.**

|                 | *1 s* | *5 s* | *10 s* | *20 s* | *30 s* |
|-----------------|-------|-------|--------|--------|--------|
| **False positives** | 896   | 166   | 49     | 18     | 12     |

**Table 6: False positives for each sliding step.**

The data in Table 5 does not show clear evidence about the benefits of shortening the window sliding step in terms of reducing the amount of data lost before detection. The average value of megabytes lost for 1 and 5 seconds of sliding window step is smaller than for 30 seconds, but less than 12%. Analysing the values for each unseen binary, all except Stop-1 and Sodinokibi-2 encrypt less data for shorter sliding steps, however, the differences are not very significant. The only exception is Sodinokibi-3, that for less than 10s window sliding step encrypts 0.27 Mbytes before detection while for 30 s step it encrypts more than 50 Mbytes.

In terms of false positives, the data in Table 6 shows a detriment when the sliding step is shortened. When the sliding step is set to 30s there is one expected alarm every 25 working days, however, a sliding step set to 1 second results in almost 3 alarms every day. As we have already discussed, high number of false alarms could disturb the users and result in the detection tool being deactivated, thus we conclude that the detriment in shortening the sliding step does not overcome the slight benefits.

## 5. Discussion and comparison with previous research

We have established a deep learning model capable of detecting ransomware while the malware is reading and writing files in a network-shared volume. The input set of features to the model describes not only the intensity of file access activity but also the number of files accessed, through what we have named as the number of short commands. These commands are recognized in encrypted traffic by an exchanged sequence of short packets and they server as the differentiator between ransomware actions and benign application actions.

Deep learning and other ML models are popular techniques in the literature on malware detection, including ransomware detection, therefore the methodological approach taken in this article is not presented as novel. However, the scenario we have described has not been significantly considered in the literature of ransomware detection. Network shared volumes are common practice in the corporate environment, but the literature on ransomware countermeasures has not deeply studied the advantages of detection based on file-sharing traffic. To the best of our knowledge, the work on [30] is the only one that takes a similar approach, but it requires unencrypted network traffic, which is accepted but not required in this article. Some of the advantages present in this scenario are present also for detection tools based on the analysis of network traffic, while some others are the result of analysing file-sharing traffic.

We have organized this discussion first on the advantages of an analysis tool based on file-sharing traffic, compared to other tools based on traffic. Following this discussion, we compare effectiveness and limitations between the deep learning mode presented in this article and the results on the previous literature. To conclude the section, we discuss the limitations and our perspectives of future work.

### 5.1 Advantages and caveats in a passive file-sharing traffic analysis scenario

Antivirus software is most often installed on the users' host. Ransomware detection tools are a specific case of antivirus tools, therefore they work in a similar manner (see for example [6] [7]). In a file-sharing environment, antivirus or ransomware detection tools can also be installed at the server, where they can monitor all files written and detect encrypted content. However, in a file-sharing scenario, a large amount of information about user actions can also be obtained from network traffic, which offers the following advantages:

- Unintrusive to the client: Antivirus software installed at the user's host can monitor CPU usage, intercept library function calls or analyse any suspect binary. However, running at the client implies being potentially intrusive to the user and consuming computer resources that could affect computer responsiveness. It is true that access operations to local files tend not to be the critical documents. They are the operating system and main programs files, which in most occasions offer only a remote desktop client environment.
- Unintrusive to the server: Although the deployment of the detection tool at the server does not directly affect the user's computer responsiveness, it has an indirect effect on it. The tool will require resources at the server and any CPU impact or disk access at the server can translate into performance degradation for all the clients. Some file servers offer simple detection techniques of large modifications at the shared content. For example, when taking a new image backup of the shared volume they can compare the size of the new compressed backup image to the previous one [48]. A compressed image much larger than the previous one can be the result of a large number of files being now encrypted (they cannot be efficiently compressed). However, this is a technique largely prone to false positives with small thresholds or very insensitive in case of large ones.
- Unintrusive to the traffic: Firewall appliances can monitor traffic, searching for anomalous patterns. They are deployed in-path, meaning that all traffic must traverse their analysis toolchain. This analysis requires processing time and in high bit rate traffic scenarios can affect file-sharing protocol performance. We suggest deploying the analysis tool off-path, monitoring a copy of the traffic between clients and servers and eliminating any effect on packet delay. When ransomware is detected, the detection tool knows the IP address of the infected host and it can be isolated by several control mechanisms. In case of having deployed a software defined network (SDN) control plane, the tool can program access rules in the network, blocking the traffic from that computer to the file server. This can also be done by programming a firewall existing in-path (without requiring the firewall to run the analysis software). If the file server is programmable, the tool can also directly remove the user access privileges at the server.
- Easy updates: A tool, such as the one described in this article, deployed off-path, is a single point to manage and its updates do not affect file-sharing service. Running a ransomware detection tool at the clients, implies an important management burden when an update is required. Running the tool at the server is simpler from a management point of view, however, being the critical server, software updates can still negatively affect the service and a failed update can be catastrophic.
- Malware resistance: When installed at the clients, the tool is vulnerable to malware infections that escalate privileges. Running the tool only at the server makes it easier to protect against malware infections. However, a tool deployed off-path is not only a single point to manage, but it can be easily protected from attacks because it does not require real network access, only monitoring a copy of the traffic. In case of requiring access to an SDN control plane it should be present only in the management network and not being accessible from the production network.

Some previous ransomware detection tools in the literature take advantage of information obtained from network traffic. Most of them cannot take advantage of above-mentioned advantages because they not only use information from network traffic but they also require to be installed at the hosts to capture other fundamental metrics [10] [12]. A shorter list of the literature takes exclusively information from network traffic. They are not based on file-sharing traffic but on DNS traffic or TCP connections to certain IPs [13] [49].

In [13] the authors use DNS traffic to recognise randomly generated domain names. Generating random DNS names is a common technique used by different malware software to locate their command and control (C&C) host without using easily-blocked well-known DNS names. This detection technique cannot be applied to ransomware strains that do not use a domain generation algorithm (DGA) and therefore are less generic than the model presented in this article.

In [14] the tool blocks the access to certain IP addresses after analysing the DNS requests. The malicious addresses must be in a blacklist that should be updated as soon as new ransomware C&C servers appear. The use of DGA by recent ransomware strains makes this task difficult and the tool cannot detect ransomwares that do not need to contact to a C&C server. Thus, the tool presented in [14] cannot detect as many strains as the tool presented in this article.

The authors of [19] analysed the network behaviour of ransomware Locky. The tool's efficiency in detecting other ransomware strains has not been verified, and it is possible that the tool could not detect ransomwares that do not contact a C&C server. They selected some TCP-level features, in addition to DNS features (domain names or DNS request failures). These features are for example the number of segments with the RST flag active or the number of hypertext transfer protocol (HTTP) POST requests sent by the user. Although they present a thorough behavioural analysis, it is focused on the ransomware Locky.

## 5.2 Comparison of effectiveness

When comparing ransomware detection proposals there are three aspects we must take into account and that limit the comparison: the ransomware strains analysed, the metrics used to evaluate performance and the reproducibility of each technique. We analyse these aspects in the following discussion.

In this article we have presented a detection technique applicable to any crypto-ransomware. In comparison, there is a subset of the literature that describes ransomware detection techniques targeting a single ransomware strain [49] [50] [51] [52]. They base the detection on searching for some special behaviour shown by the selected strain and therefore they are hardly generalisable to other ransomware families. Those techniques are more effective for the specific ransomware they target than any generic approach, but they have not been compared with other families and in most cases, they cannot be applied because the mechanisms they search for are not present in different strains. We will not elaborate any further in the comparison with these proposals because they do not apply to a realistic production environment where any ransomware strain can infect a host.

The rest of the literature takes a set of ransomware binaries from some repositories. They can be organised in strains or families [11] [13] [17] or they can be a large set that simply results from searching in a database using ransomware-related keywords [12] [16] [53]. In recent years we have witnessed an important increment in new ransomware appearances. Depending on the age of the ransomware detection proposal, it has been tested with a different set of strains. Algorithms described in decade-old papers can be designed to detect ransomware binaries that we cannot run nowadays because their C&C servers are down or the DNS names they try to resolve are blocked. Comparison with these papers is problematic because their results cannot be generalised to new ransomware strains and new methods cannot be tested with the families present at their time.

The metrics used to evaluate performance of the detection algorithm are usually centred on the capability to detect the ransomware and on its erroneous classification of benign applications as ransomware. These two metrics relate to the true positives and the false positives in a binary classification problem. Ransomware detection can be measured on the basis of whether a ransomware binary is recognised as malware or not; it can also be measured using the time the algorithm requires watching ransomware actions to classify it as malware, or it can be measured as the amount of data lost before detection. Some of these metrics depend on the environment (computing power, disk files distributions, simultaneous user actions) and there is no consensus on a single metric whose value could be compared in the same circumstances in future works. A similar problem arises when discussing false positive classification, where critical parameters are for example the type or number of benign applications being run, or the user activity pattern. Due to the different metric definitions, objective comparability in numeric results is not reliable and we will have to consider what each metric is really measuring and in which context.

Finally, some of these handicaps can be solved if previous proposals can be tested against new data. This requires a clear definition of the algorithms, making their implementation reproducible. A second option is publishing all the data used, so new algorithms can be compared in the same scenarios. It is not the objective of this section to enumerate the literature on ransomware detection which lacks in comparability due to unclear algorithm definitions or unpublished data. We just tried to make our best contribution to facilitate future comparability by making available both the dataset used in this article and the neural network that we obtained after training.

Taking all these aspects into account, we proceed to discuss the results in previous papers having a number of ransomware strains similar to those in this study. The tool we have described requires only a single 30 s sample from network traffic to detect more than 99% of the ransomware samples we tested (150 traffic traces from more than 30 strains appearing in a period of five years). In the remaining cases (less than 20%), 2 samples (60 s) are required. We have divided the existing tools into two groups depending on where they have to be installed.

### 5.2.1 Locally installed tools

These tools must be installed in each host that has access to the shared files to provide effective protection. They need some features that can only be obtained from the local machine, such as certain use of library calls, disk-access actions, the identification of the processes running, the cryptographic primitives used, or the file-contents being written. However, these tools can impact a machine's resource utilization, and they must be kept updated on every machine in the network.

**N. Scaife et al** [11] measured disk-access actions from each running process. Results of testing the described tool with ransomware binaries from 24 different strains revealed 100% detection rate and 1/30 applications causing a false positive. On average, 10 files were encrypted before malware detection, although it depends on the file-size and on the ransomware strain. In the case of our tool, 51 files were lost on average in a worst-case scenario of 7 unseen ransomware binaries. Our ransomware dataset was collected between 2015 and 2020, whereas the tool in [11] was trained and evaluated using ransomware binaries that appeared up to 2016 and it has not been tested with more recent ransomware strains such as WannaCry, notPetya, or Phobos.

**D. Sgandurra et al** [17] based ransomware detection on the features extracted from a dynamic and static analysis and employed a supervised regularized logistic regression ML algorithm. The tests were performed using ransomware binaries from 11 different strains that appeared before 2016 and 942 manually executed user applications. They achieved a detection rate of 96.3% and false positive rate of 1.6%. We have presented a solution that improves the results in terms of detection and false positive rate.

**H. Zuhair et al** [54] presented a hybrid ML model combining DTs and naïve Bayes models. The results of their model using 14 ransomware strains and 500 different applications are 97% accurate. Our tool improves the results in terms of accuracy using a higher number of ransomware strains for evaluation.

**S. Mehnaz et al** [44] used the random forest algorithm for the classification, using file-access primitives as input features. They achieved a detection rate of 100% and a false positive rate of 0.9%. The proposal can detect ransomware before the encryption of 10 files. This solution, however, introduces significant latency in I/O operations and overloads the user machine. With our tool, the results concerning false positive rate (0.004%) were improved even when the information about the client operations cannot be extracted from the encrypted network traffic.

**A. Continella et al** [7] offered file recovery after a ransomware attack in addition to its detection. The detection and false positive rates were 97.7% and 0.038% respectively. Similar to [11], the ransomware dataset used covered only the ransomware strains appeared until 2016.

**R. Moussaileb et al** [22] analysed the file-system traversal actions performed by the ransomware to locate the files to be encrypted. They compared the classification results using a random forest, DTs, and k-nearest neighbours algorithms. DTs provided the best accuracy with a detection rate of 99.35% and a false positive rate of less than 1%. It bases its detection on a certain ransomware behaviour that is not presented in all strains; therefore, some ransomware strains could evade the detection.

### 5.2.2 Network based tools

These tools do not require their installation in the user machine because they do not need any information monitored locally. The main advantages compared to the locally installed group is that they do not consume resources in the user machine, and they cannot be deactivated by ransomware that escalates privileges in the infected host. However, they do not have access to local host information, which can hamper their capabilities.

**K. C. Roy et al** [25] achieved ransomware detection by analysing the log files sent by the user machines to a server. The solution extracts some features from these logs and uses them as inputs to a bidirectional long short-term memory NN, which is a type of recurrent NN. The proposed solution achieves 99.87% detection accuracy using 17 different ransomware strains and 0 false positives. No experiments with unseen samples are performed in this study; therefore, the solution could have problems in detecting ransomware strains that are not present in the training phase.

**S. Chadha et al** [18] detected whether the names in DNS requests are generated by a domain generation algorithm or not. They compare supervised and unsupervised ML algorithms, obtaining an optimum configuration with a detection rate of 85% and false positive rate less than 10%. Ransomware detection based on DNS traffic fails for ransomware strains that do not need to contact an external server or they establish the connection after the encryption process [37].

**A. O. Almashhadani et al** [19], similar to [18], detected ransomware by analysing the DNS requests and training using different ML algorithms, including DTs and TEs. They achieved a detection rate of 97.8% and a false positive rate of 0.04%. It has the same limitations as [18] because it uses the same features for detection.

**D. Morato et al** [30] described a tool that is most similar to the one presented in this paper. It detects ransomware by analysing traffic from SMBv2. It achieves 100% detection rate and 1 out of 10 billion false positives. However, owing to the features it extracts from the traffic, it is not applicable to an SMBv3 scenario, wherein the file-access commands are encrypted. We have generalized the scenario with a new tool, capable of ransomware detection in encrypted file-sharing scenarios. This encryption limits the available information in network traffic, causing the false positive rate to increase compared to [30]. However, the tool can detect the unseen ransomware binaries downloaded from [37], losing an average of only 37 MB data before detection.

*5.3 Limitations and future work*

We have described a ransomware detection technique based on file-sharing traffic from Microsoft Windows desktop computers that are the potential target of ransomware infections. We do not consider mobile operating systems, where the file-sharing scenario is unlikely. SMB is the prevalent protocol in networks where the hosts use Windows operating system, therefore we have evaluated the effectiveness of our proposal using SMBv3 traffic. The messages from this protocol are encrypted, therefore the message command, data or metadata cannot be analysed. This has required defining input features to the ML models that do not need data that is protocol-specific. We believe that the result is a very protocol-agnostic detection procedure and we pretend to validate the results using other file-sharing protocols. In an environment with a mixture of Microsoft Windows and UNIX operating systems, the file-sharing protocol of choice could be the network file sharing (NFS) protocol. We have evaluated our neural network model with network traffic obtained from running the unseen samples in an NFS environment and the results were similar to those obtained from SMB in terms of detection capability and lost data before detection. Although NFS used from windows desktops is not a frequent network scenario, we pretend to validate the results obtained from an SMB scenario by running the new appearing ransomware samples in a mixture of SMB and NFS scenarios. This would prove the generalization capabilities of presented model.

Finally, we have described a static solution, trained with ransomware strains from 2015 up to 2020 that we expect to be valid when new families appear, however, nothing in the analysis supports this affirmation. We pretend to create better adaptive training methodologies where new ransomware strains can be incorporated into the ML model and evaluate the improvement or deterioration of the results.

# 6. Conclusions

In this paper, we described and validated a tool that can detect more than 70 ransomware binaries acting in a file-sharing scenario using encrypted protocols. The tool works with a copy of the traffic; therefore, it does not affect user activity. The detection is based on the bytes read and written from the file server and the control commands performed by the user on the files.

We explained the feature extraction and sample reduction processes before the selection of the best ML model. The comparison between decision trees, tree ensembles, and neural networks reveals that neural networks provide the best results using 3 hidden layers of neurons. The validation reveals that the model has a false positive rate of 0.004% with more than 1,400 h of real user traffic. It can detect all ransomware binaries used in the training phase in an average time of 30.2 s. It detects 100% of a set of ransomware binaries not used in the training phase, losing only an average of 114 megabytes of user data before detection.

The time window length is a tuneable parameter in the feature extraction process, and it must be configured depending on the scenario and user and server characteristics. For the scenario presented in this paper, the best trade-off in the results was obtained with a time window of 30 s. With larger values of T, the ransomware encrypts a significant number of bytes (more than 100 MB on average). Shorter time windows result in a higher number of false positives, which could annoy the network administrator and make the tool useless. Finally, we compared the model in this paper with other ransomware detection tools in the literature, Despite the novelty of the scenario that hinders comparison, the tool improves most of the results found in the literature.

## Acknowledgments

This work was supported by the Spanish State Research Agency with project PID2019-104451RB-C22/AEI/10.13039/501100011033.